\documentclass[11pt]{article}

\usepackage[all,arc,curve,frame]{xy}

\usepackage{epsfig}

\usepackage{framed}

\usepackage[english]{babel}
\usepackage[utf8]{inputenc}

\usepackage{amsmath}
\usepackage{amssymb}
\usepackage{theorem}

\newtheorem{theorem}{Theorem}[section]
\newtheorem{lemma}[theorem]{Lemma}
\newtheorem{corolary}[theorem]{Corolary}

\usepackage{listings}
\usepackage{alltt}

\usepackage{graphicx}

\usepackage{comment}

\newcommand{\forPS}{\epsfig{file=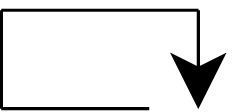,width=5ex}}

\newcommand{\for}{\raisebox{-1ex}{\forPS} }

\def\contador#1{$\xrightarrow{~~~~~#1~~~~~}$}

\newcommand\startBloc{%
  \hspace{-1ex}\vrule\begin{minipage}{\linewidth}
\begin{list}{}{\leftmargin=3em \labelsep=0pt \labelwidth=0pt
      \rightmargin=0pt \topsep=0cm \partopsep=0cm \parsep=0pt
	  \itemsep=0pt \labelwidth=0pt
	  \itemindent=0pt
	  \listparindent=0pt
	  \parskip=0pt
	  }
	  \item[]~\\
}

\newcommand\stopBloc{
 \end{list}
 \rule{4ex}{0.5pt}
 \end{minipage}
 ~
}

\newcommand\stopBlocNL{
 \end{list}
 \end{minipage}
 ~
}

\newcommand{\jO}{\,\vee\,}
\newcommand{\jY}{\,\wedge\,}

\newcommand{\entonces}{\Rightarrow}

\newcommand{\es}{\leftarrow}

\title{Two Mutual Exclusion Algorithms\\ for Shared Memory}
\author{Jordi Bataller Mascarell\\
Departament de Sistemes Informàtics i Computació\\
Universitat Politècnica de València (SPAIN)\\
bataller@dsic.upv.es} 
\date{\today}

\begin{document}


\maketitle

\begin{abstract}

  In this  paper, we  introduce two algorithms  that solve  the mutual
  exclusion problem for concurrent  processes that communicate through
  shared  variables,  \cite{Dijkstra:Solution:1965}.   Our  algorithms
  guarantee that  any process  trying to  enter the  critical section,
  eventually, does enter it.  They are formally proven to be correct.

  The first algorithm  uses a special coordinator process  in order to
  ensure equal chances to processes waiting for the critical section.
  In the  second algorithm, with  no coordinator, the  process exiting
  the  critical section  is in  charge to  fairly elect  the following
  one. In the case that no process is waiting, the turn is marked {\em
  free} and will be determined by future waiting processes.
  The type  of shared variables used are a
  $turn$  variable, readable  and  writable by  all  processes; and  a
  $flag$  array,  readable by  all  with  $flag[i]$ writable  only  by
  process $i$. There is a version of the first algorithm where
  no writable by all variable is used.

The bibliography reviewed for this paper is
\cite{Lynch:Distributed:1996} and \cite{Gries:Logical:1993},
all the rest is original work.

\end{abstract}



%

\newpage
\section{Asymmetric Algorithm}

This algorithm uses  a coordinator process in charge  of assigning the
right to access the critical section to processes trying to enter it.
The algorithms are written using
the notation described in \cite{Bataller:Visual:2011}.

Shared variables
\begin{itemize}
\item Processes $\in \{p_0, ... , p_{N-1}\}$, plus the coordinator.

\item  $turn  \in  \{0,  ... ,  N-1,  \text{THINKING}  \}$,  initially
  THINKING, readable and writable by all.

\item $flag[i] \in \{\text{REMAINDER, WAITING}\}, i \in \{0,...,N-1\}$.
Initially,  $(\forall   i  |   0  \leq   i  \leq   N-1  :   flag[i]  =
\text{REMAINDER})$.
$flag[i]$ is readable by all and writable by $p_i$.
\end{itemize}

A process  $p_i$ willing to enter  the critical section announces  this by
setting
$flag[i]$ to WAITING, and then waits until $turn$ is $i$.
On exit, it reverts $flag[i]$ to REMAINDER and sets $turn$ to
THINKING, so the coordinator knows the process is done
and turn must be updated.

\begin{framed}
  {\bf Algorithm for process $p_i$} \\
  
$flag[i] \es$ WAITING \\
\for $turn \neq i$ \\

CRITICAL SECTION \\

$flag[i] \es$ REMAINDER \\
$turn \es \text{THINKING}$
\end{framed}

The coordinator performs  an infinite loop where  it serially inspects
the state  of processes.  When a  processes is  found waiting  for the
critical section,  the coordinator grants  the access to it;  and then
waits  until that  process leaves  the section --signaled by
setting $turn$ to THINKING.

\begin{framed}
{\bf Algorithm for the coordinator process} \\

$p \es 0$ \\
\for $\infty$ \\
\startBloc
$flag[p] = \text{WAITING} \entonces$ \\
 \startBloc
 $turn \es p$ \\
 \for $turn \neq$ THINKING 
 \stopBloc

 $p \es (p+1)$ mod $N$
\stopBloc
\end{framed}

Although the above algorithms  are written mathematically, problems on
concurrency are difficult  to reason about.  It is  much more suitable
to  define algorithms  by means  of finite  state automata  with edges
representing events  --transitions between  two states.  An  event may
have a precondition, written  ``{\em precondition} $\Rightarrow$''; or
an  effect,  written  ``$\Rightarrow$  {\em  effect}''.   This  is  a
practical simplification of the formalism in \cite{Lynch:Distributed:1996}.

\begin{framed}
{\bf Automaton for process $p_i$.}

\centerline{
\xymatrix{
  \fbox{~1~} \ar[d]^{\entonces flag[i]=\text{WAITING}}\\
  \fbox{~2~} \ar[d]^{turn=i \entonces } \\
  \fbox{critical section} \ar[d]^{\entonces flag[i]=\text{REMAINDER}}\\
  \fbox{~3~} \ar[d]^{\entonces turn = \text{THINKING}} \\
  \fbox{~4~} 
} 
} 
\end{framed}

\begin{framed}
{\bf Automaton for the coordinator process.}

\centerline{
\xymatrix{
  \fbox{~1~} \ar[d]^{\entonces p=0}\\
  \fbox{~2~} \ar[d]^{flag[p]\neq\text{WAITING}\entonces}
     \ar[rrr]^{flag[p]=\text{WAITING}\entonces} &&&
     \fbox{~2.1~} \ar[rr]^{\entonces turn=p} &&
     \fbox{~2.2~} \ar@/^1pc/[dlllll]^{turn=\text{THINKING}\entonces}\\
  \fbox{3} \ar@/^2pc/[u]^{\entonces inc(p)\text{mod} N} 
} 
} 
\end{framed}

\begin{theorem}
\label{theorem:MutualExclusion}
Mutual exclusion.

No  two  processes can  be  at  the same  time  in  the {\em  critical
  section} state.
\end{theorem}

{\bf Proof of theorem \ref{theorem:MutualExclusion}}

Suppose, for a contradiction, that there exists
an execution where process $b$ is the first one to enter the critical
section while another one, $a$, is still inside.
This timeline shows the situation.

~\\
\centerline{\epsfig{file=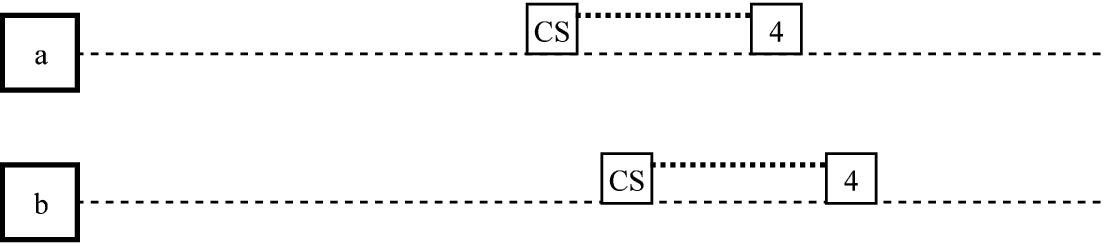,width=0.8\linewidth}}



If $b$ has to  enter the critical section while $a$  is in, there must
be another  process, say $c$, to  change $turn$ to THIKING  before $a$
does  (within  the lapse  $[\text{CS}_a,  4_a[$),  so the  coordinator
    scapes from 2.2 and can reach $2.2$ again to let $b$ in.  But that
    means  that $c$  is in  the critical  section overlapped  with $a$
    before $b$ overlaps with $a$.

~\\
\centerline{\epsfig{file=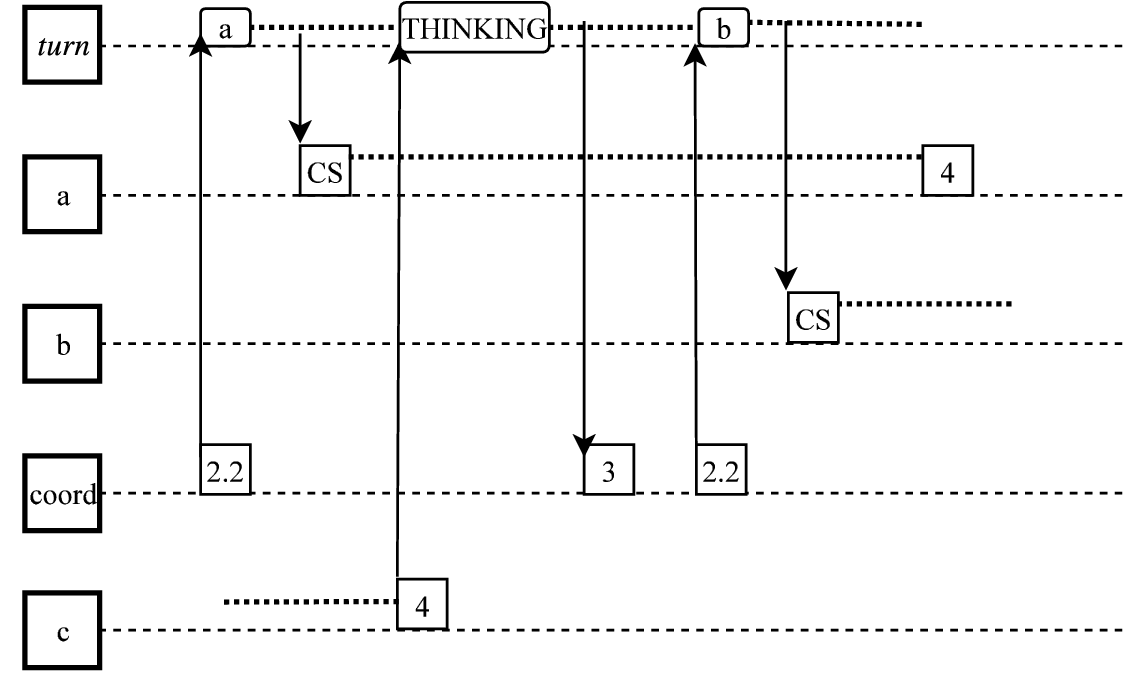,width=0.8\linewidth}}

    This is  a contradiction,  as we
    assumed that $b$ is the first one breaching the critical section.
$\Box$

\begin{lemma}
\label{lemma:ProcessWithTurnEnters}

If the turn is assigned to a process, then, it will enter the critical
section.
\end{lemma}

{\bf Proof of lemma \ref{lemma:ProcessWithTurnEnters}}

If process $a$ is in a state $2$ and turn is assigned to it --as the
coordinator reaches 2.2;  only $a$ may enter the  critical section (by
the condition  $turn=i\Rightarrow$) since  the coordinator will  be in
state 2.2 while $turn\neq \text{THINKING}$.
In this point, there is no other process in the critical section (or $a$
could violate mutual exclusion) able to change $turn$ to THINKING.

A  second  case is  when  process  $a$  leaves the  critical  section.
Because  $a$   sets  $flag[i]$  to  REMAINDER   before  signaling  the
coordinator to choose the following  process for the critical section,
with $turn=\text{THINKING}$,  $a$ can't  reach state  4 with  the turn
reasigned to it.  $\Box$

\begin{theorem}
\label{theorem:ProcessWillEnter}
A proceess waiting in state 2 will enter the critical section.
\end{theorem}

{\bf Proof of theorem \ref{theorem:ProcessWillEnter}}

The coordinator asigns turns in a round-robin fashion between
processes with $flag[i]=\text{WAITING}$ --in the loop
$2, 2.1, 2.2, 3$ using counter $p$ .

Once a process $a$ is waiting in 2, 
processes  with  number   $[a+1,...,N-1,0,...,a-1]$  could  enter  the
critical section before $a$. If one such process is not waiting,
$p$ increases towards $a$; if it is waiting, it will enter the critical
section and, at exit, $p$ is equally increased towards $a$.
Finally, $p=a$ and $a$ will enter the critical section.
$\Box$

\section{Asymmetric Algorithm with one-writer/multiple-reader variables}

A new  version of our  previous algorithm can  be devised in  order to
have  the $turn$  variable  to  be writable  only  by the  coordinator
process.  This is at the expense of making the exiting process, $p_i$,
to wait  until the coordinator  changes $turn$ to  something different
from  $i$. Otherwise,  $p_i$ would  be able  to re-enter  the critical
section using its previous $turn$. And worse, the coordinator
could change $turn$ to a different waiting process and the
mutual exclusion would be broken.

Shared variables
\begin{itemize}
\item Processes $\in \{p_0, ... , p_{N-1}\}$, plus the coordinator.

\item  $turn  \in  \{0,  ... ,  N-1,  \text{THINKING}  \}$,  initially
  THINKING, writable only by the coordinator process, readable by all.

\item $flag[i] \in \{\text{REMAINDER, WAITING}\}, i \in \{0,...,N-1\}$.
Initially,  $(\forall   i  |   0  \leq   i  \leq   N-1  :   flag[i]  =
\text{REMAINDER})$.
$flag[i]$ is readable by all and writable by $p_i$.
\end{itemize}

\begin{framed}
  {\bf Algorithm for process $p_i$} \\
$flag[i] \es$ WAITING \\
\for $turn \neq i$ \\

CRITICAL SECTION \\

$flag[i] \es$ REMAINDER \\
\for $turn \neq$ i
\end{framed}

\newpage
\begin{framed}
{\bf Algorithm for the coordinator process} \\
$p \es 0$ \\
\for $\infty$ \\
\startBloc
$flag[p] = \text{WAITING} \entonces$ \\
 \startBloc
 $turn \es p$ \\
 \for $flag[p] \neq$ REMAINDER \\
 $turn \es \text{THINKING}$ 
 \stopBloc
 $p \es (p+1)$ mod $N$
\stopBloc
\end{framed}

\section{Symmetric Algorithm}

This algorithm has  no special coordinator process, but,  in some way,
it integrates its task.  When a process exits the critical section, it
fairly  decides  which waiting  process  is  granted the  next  access
--setting variable  $turn$.  If no  one is  waiting, $turn$ is  set to
FREE, same as at the beginnig.   If the critical section is found free,
incoming  processes will  elect  one  of them  to  enter the  critical
section.

Shared variables
\begin{itemize}
\item Processes $\in \{p_0, ... , p_{N-1}\}$

\item  $turn  \in \{0,  ...  ,  N-1, \text{THINKING},  \text{FREE}\}$,
  initially FREE, readable and writable by all.

\item $flag[i] \in \{\text{REMAINDER, WAITING, CANDIDATE}\}, i \in 
\{0,...,N-1\}$.
Initially,   $(\forall   i  |   0   \leq   i   \leq  N-1   :   flag[i]
= \text{REMAINDER})$.
$flag[i]$ is readable by all and writable by $p_i$.

\end{itemize}

\newpage
\begin{framed}
{\bf Algorithm for process $p_i$, entry part} 

$flag[i] \es$ WAITING \\
\for $turn = \text{THINKING}$ \\

$turn = \text{FREE} \entonces$ \\
 \startBloc
 $flag[i] \es$ CANDIDATE \\
 \for $turn = \text{FREE} \jY flag[i]=\text{CANDIDATE}$ \\
  \startBloc
  $\text{nCandidates} \es 0$ \\
  $\text{minCandidate} \es -1$ \\

  \for $N-1$ \contador{j} $0$ \\
   \startBloc
     $flag[j] = \text{CANDIDATE} \entonces$ \\
     \startBloc
       $\text{nCandidates} \es \text{nCandidates}+1$ \\
       $\text{minCandidate} \es j$
     \stopBloc
   \stopBloc
  ~ \\
   $turn=\text{FREE}\jY\text{nCandidates} = 1 \entonces$ \\
     \startBloc
       $turn \es i$
   \stopBloc
   $turn\neq\text{FREE}\jO \text{minCandidate} < i \entonces$ \\
          \startBloc
              $flag[i] \es$ WAITING 
          \stopBloc
  \stopBloc 
  ~ \\
   $flag[i] \es$ WAITING 
 \stopBloc

\for $turn \neq i$ \\

CRITICAL SECTION 
\end{framed}

\newpage
\begin{framed}
{\bf Algorithm for process $p_i$, exit part} \\

CRITICAL SECTION \\

$turn \es$ THINKING \\
$flag[i] \es$ REMAINDER \\
$nextTurn \es$ THINKING \\
  \for $nexTurn=\text{THINKING} \jY$ $i+1$ \contador{j} $N-1$, 0 \contador{j} $i-1$ \\
 \startBloc
   $flag[j] \neq \text{REMAINDER} \entonces$ \\
     \startBloc
       $nextTurn \es j$
   \stopBloc
 \stopBloc

   $nextTurn = \text{THINKING} \entonces$ \\
     \startBloc
       $turn \es $ FREE
   \stopBloc
   $nextTurn \neq \text{THINKING} \entonces$ \\
     \startBloc
       $turn \es nextTurn$
   \stopBloc
\end{framed}

\newpage
\thispagestyle{empty}
\enlargethispage{12\baselineskip}
This is the automaton version of the algorithm. \\

\hrule
{\bf Entry part}
~\\
\centerline{\epsfig{file=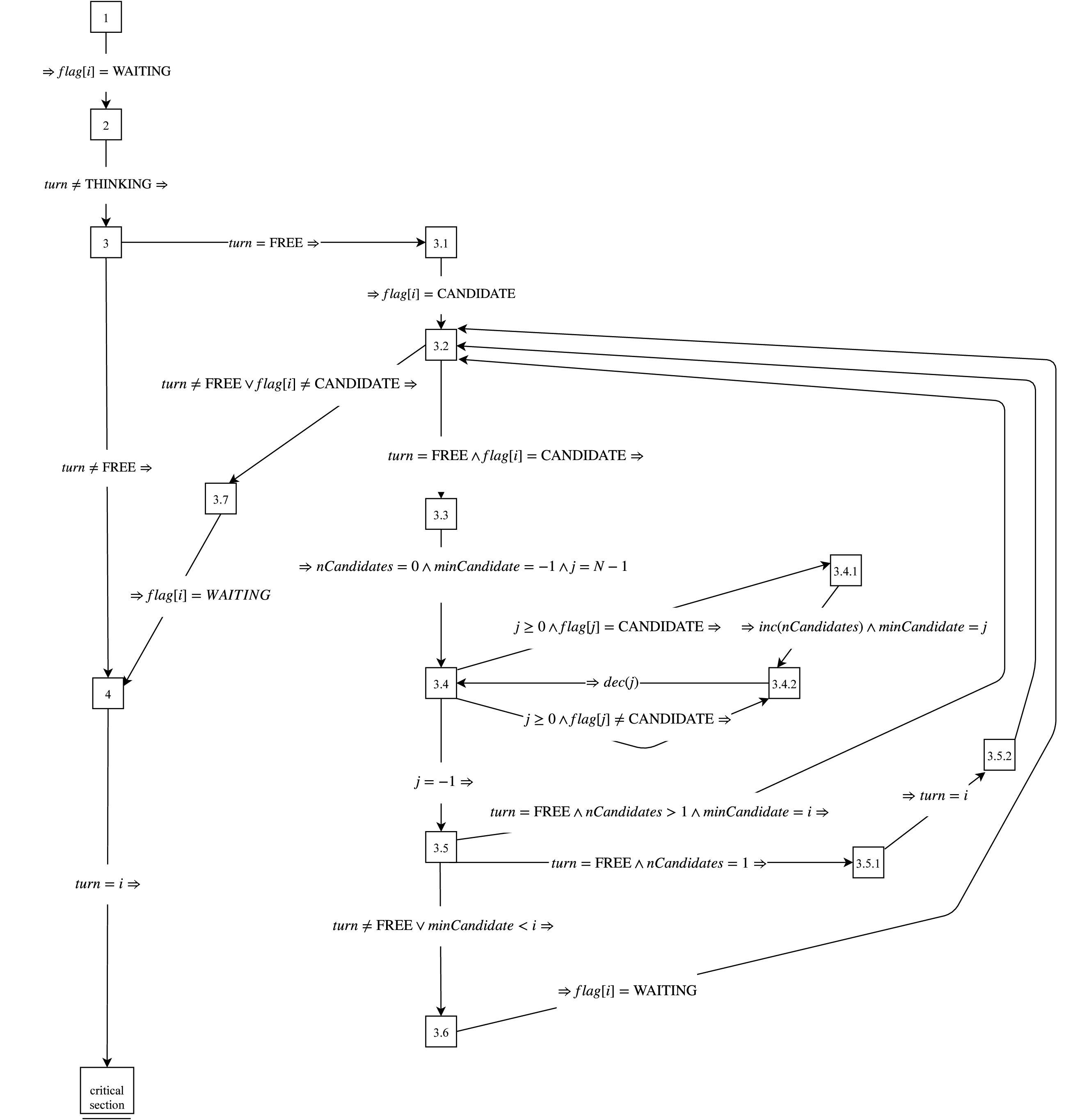,width=1.5\linewidth}}


\newpage
\hrule
{\bf Exit part} \\

\centerline{\epsfig{file=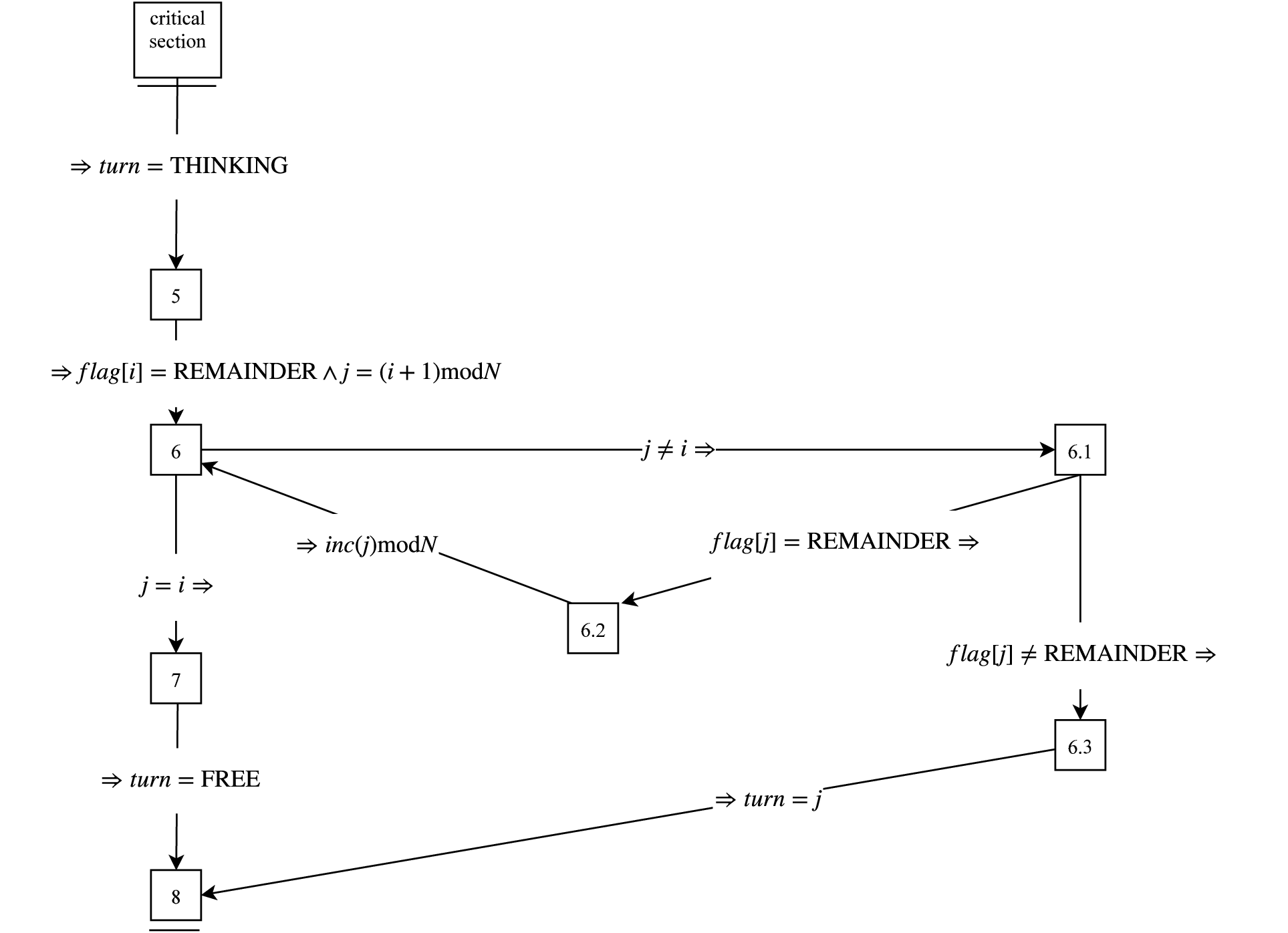,width=1.3\linewidth}}
~\\
\hrule

\newpage

\begin{lemma}
\label{lemma:NoTwoIn351}

No two processes are in state 3.5.1 simultaneously.
\end{lemma}

{\bf Proof of lemma \ref{lemma:NoTwoIn351}}
Suppose for a contradiction that two processes $a$ and $b$
are in state $3.5.1$, being $a$ the first to arrive.

The number of
processes  with $flag[i]=\text{CANDIDATE}$  is  computed  in the  loop
$3.4, 3.4.1, 3.4.2$.
When $a$ arrives at $3.5.1$, $nCandidates=1$.
That means that
$b$ has not reached state $3.2$ (its flag is not CANDIDATE)
when $a$ inspects $flag[b]$.

Now, by assumption, $a$ remains in $3.5.1$ until $b$ arrives.
That implies that $flag[a]=\text{CANDIDATE}$ during that lapse.
Eventually, $b$ passes $3.2$, marking itself as candidate
and, when it ends the $3.4$ loop, $nCandidates$ for $b$ must be at least
2 (itself and $a$).

~

\centerline{\epsfig{file=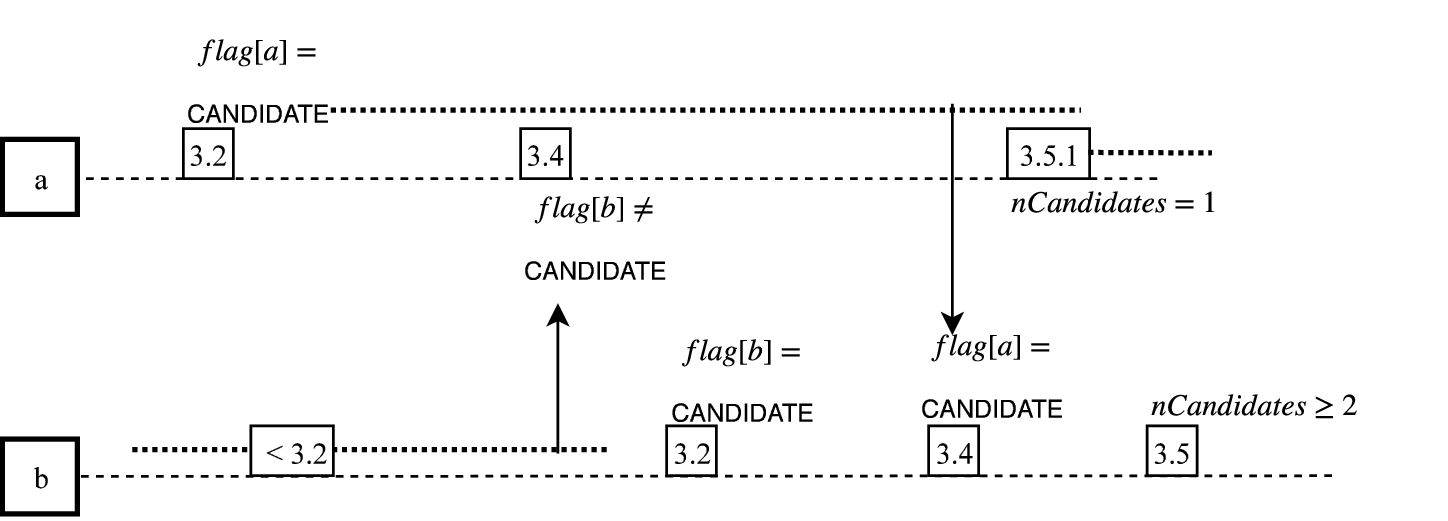,width=\linewidth}}

Thus, $b$ can't enter $3.5.1$ as the condition
to do so fails. This contradicts the assumption that both $a$ and $b$
can be in $3.5.1$ simultaneously.
$\Box$

We  say that  a  turn assignment  (like  the one  when  state 3.5.2  is
reached) is an {\em enabling turn assignment} iff that assignment is the
latest before the process granted by it enters the critical
section.

The following  timeline shows an  enabling turn assignment  for process
$a$.
In the marked lapse, $turn$ is not changed.
~\\
\centerline{\epsfig{file=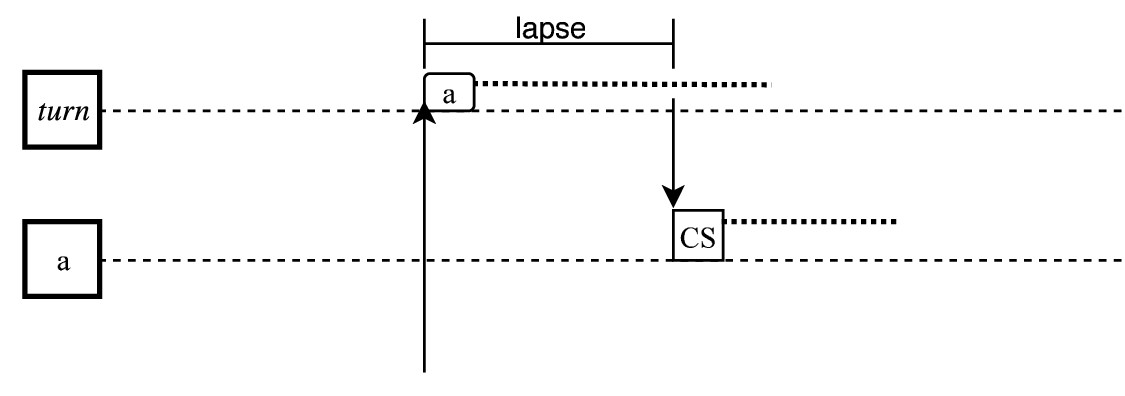,width=0.7\linewidth}}

\newpage
\begin{theorem}
\label{theorem:MutualExclusion2}
Mutual exclusion.

No two processes can be at the same time in the {\em extended
critical section} --states in $[\text{critical section}, 8[$.

\end{theorem}

\begin{corolary}
(of theorem \ref{theorem:MutualExclusion2})

Two processes can't be simultaneously in the critical section.
\end{corolary}

{\bf Proof of theorem \ref{theorem:MutualExclusion2}}

Suppose,  for  a contradiction,  that  there  exists an  execution  where
process $b$  is the first one  to enter the extended  critical section
(states $[\text{critical  section}, ...,  8[$) while another  one, say
$a$, is still in. This is:

~

\centerline{\epsfig{file=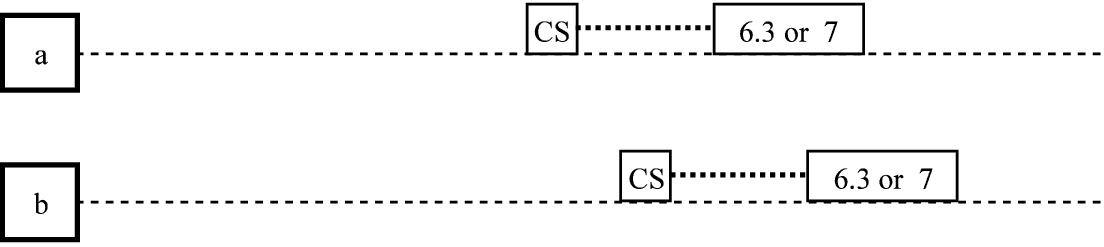,width=0.7\linewidth}}

Now, let's consider  the cases where the  enabling turn assignments
of $a$'s and $b$'s can happen.

\begin{itemize}
\item Case 1. The enabling assignments for both processes $a$ and $b$ occur at 3.5.2.

By lemma
\ref{lemma:NoTwoIn351}, when $a$ is in $3.5.1$, $b$ is not.
From  $3.5.1$, $a$  reaches  the extended  critical  section, and,  by
assumption, it remains  {\em alone} within it (until  $b$ also arrives
at).   Thus, during  this  lapse,  $turn$  is either  $a$  or
THINKING, never FREE:

~

\centerline{\epsfig{file=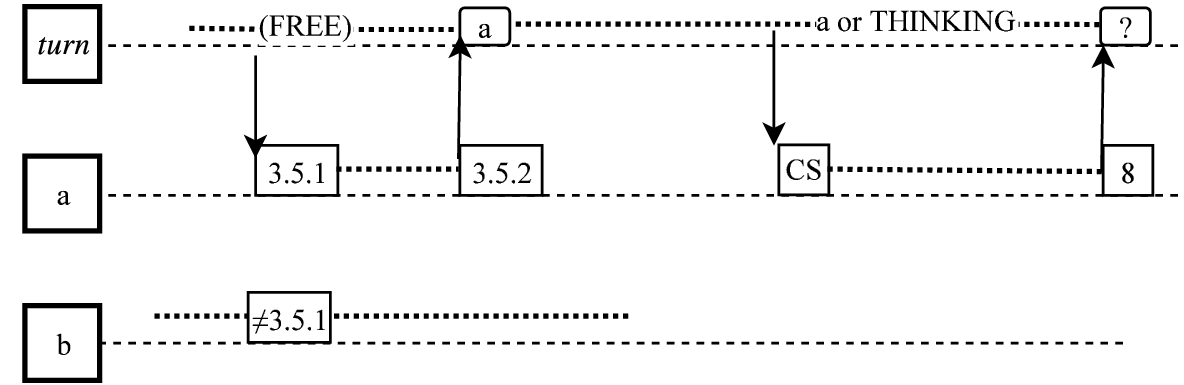,width=0.7\linewidth}}

As a consequence, $b$ can't access state $3.5.1$ (nor $3.5.2$) to
perform  its enabling  assignment,  since $turn\neq\text{FREE}$.   This
contradicts this case.

\item Case 2. No matter which is the enabling assignment for $a$,
the enablig assignment for process $b$ is performed by
a third process $c$ going from state $6.3$ to $8$.
This would be the situation.

~

\centerline{\epsfig{file=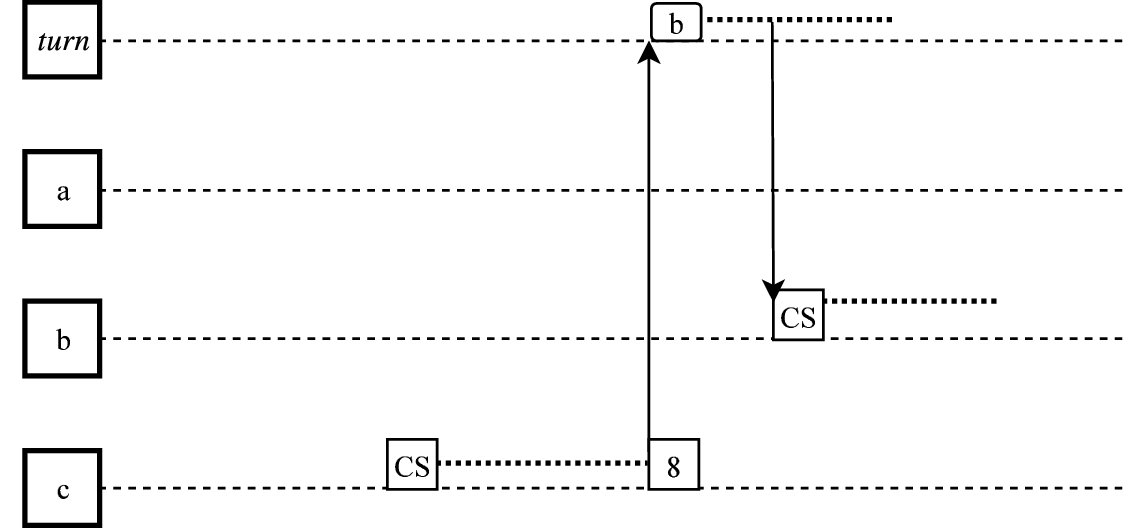,width=0.7\linewidth}}

By the contradicting  assumption, $a$ must enter  the critical section
before $b$ (and remain there until it overlaps with $\text{CS}_b$).
In this case, $\text{CS}_a$ must be before $8_c$ as well, because $turn=b$
is set at that moment and it is fixed until $\text{CS}_b$.
This implies that $a$ must be overlapped in the critical section with
process $c$. This  contradicts the asumption that $a$ and  $b$ are the
first processes that break the mutual exclusion.

\item Case 3 (last).
The enabling assignment for process $a$ is done by a third process, say
$c$; and the enabling asigment for $b$ occurs at state $3.5.2$.
This is the outline now:

~

\centerline{\epsfig{file=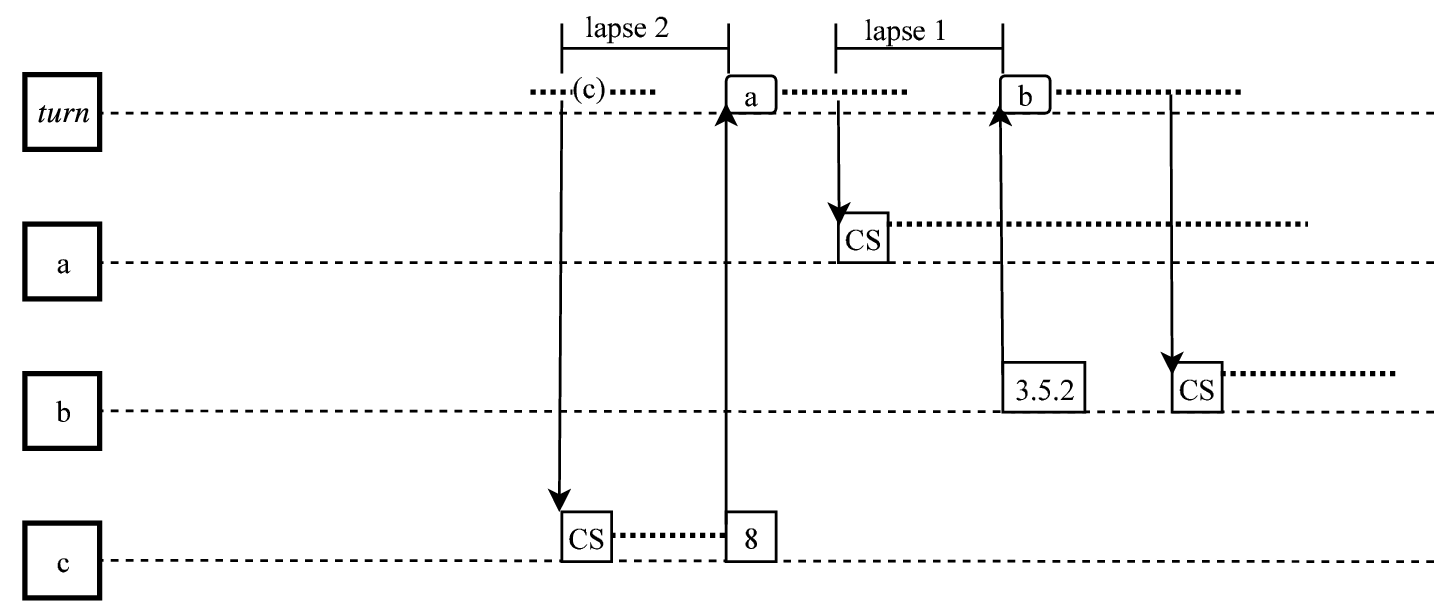,width=\linewidth}}

Let's consider when $b$ arrived at state $3.5.1$. Because
this needs
$turn=\text{FREE}$, a process must assignt FREE to $turn$
as it reaches its state $8$.

The assignment can't be in $[\text{CS}_a, 3.5.2_b]$ (lapse 1), because
that process would overlap its critical section with $a$'s.
Neither in $[8_c,  \text{CS}_a]$ because $turn=a$ is fixed  for $a$ to
enter the critical section. Neither in $[\text{CS}_c, 8_c]$ (lapse 2),
because that overlaps with $c$'s critical section.
Therefore, $b$ is in state $3.5.1$ before $\text{CS}_c$. This is:

~

\centerline{\epsfig{file=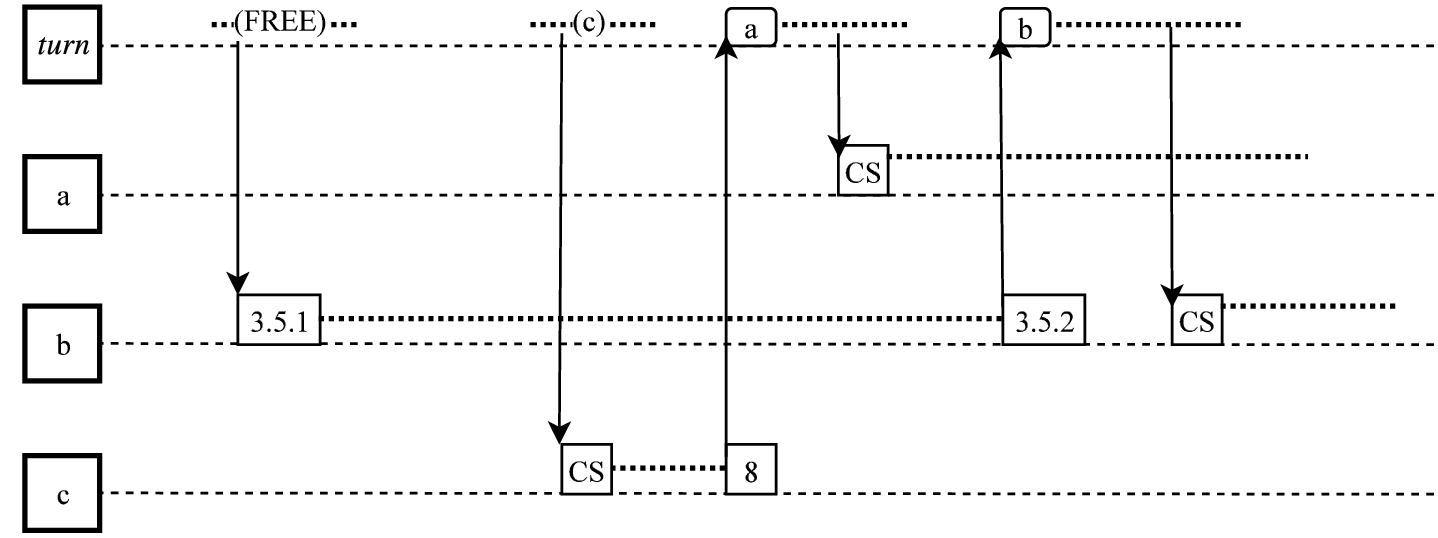,width=\linewidth}}

Now let's consider how $turn$ is set to $c$ in $[3.5.1_b, \text{CS}_c]$.
In this interval, by lemma
\ref{lemma:NoTwoIn351}, $c$ cannot do
a turn asigment to itself from state $3.5.1$.
Hence, a process, say $j$, does this assignment in its state $8$.
When did $j$ enter its critical section? If later than $3.5.1_b$,
then  again,  another  process  in $8$  should  have  performed  $j$'s
enabling assignment after $3.5.1_b$.  But there may exist only
a finite number of proceses 
entering its critical section
in $[3.5.1_b, \text{CS}_c]$ and enabling the following one until $j$ and
then $c$. Let's suppose it is process $k$ the one which
is  in state  $8$  after  $3.5.1_b$ but  enters  its critical  section
before.
This is depicted in the following timeline.

~

\centerline{\epsfig{file=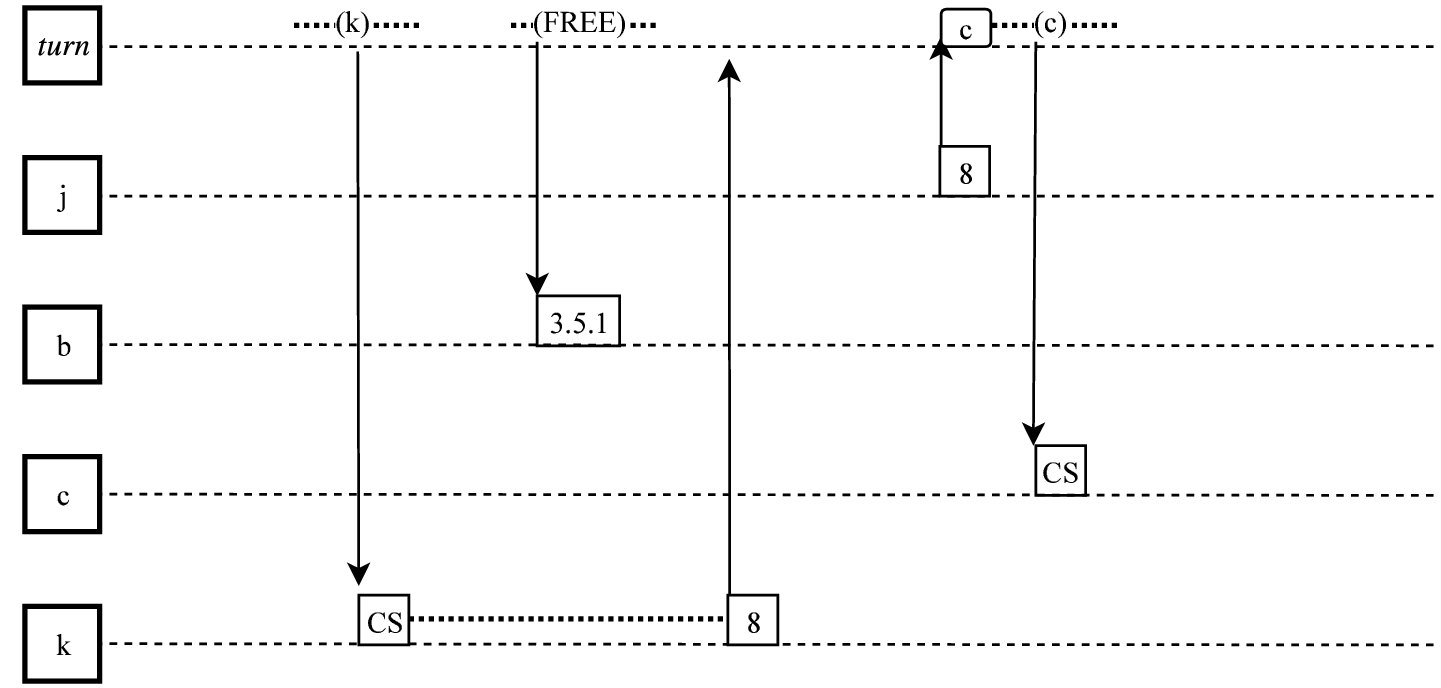,width=\linewidth}}

Finally, we reach the desired contradiction:
$turn$ can't be changed to FREE 
in $[\text{CS}_k, 8_k]$ because there cannot exist another
process in the critical section overlapped with $k$ 
--$a$ and $b$ are the first ones to overlap.

\end{itemize}

$\Box$

\begin{lemma}
\label{lemma:AssignmentFixed}
If the turn is assigned to a process, then, 
no other one will change this assignment
before it enters the critical section.
\end{lemma}

{\bf Proof of lemma \ref{lemma:AssignmentFixed}}

If a process $a$ assigns the turn to itself in state $3.5.2$,
the assignment won't be changed
(before $a$ enters the critical section):

\begin{itemize}
\item Not at state $8$. Otherwise, this would imply
that a process is in 
$[\text{critical section}, 8[$
and $a$ could use its turn to break
theorem \ref{theorem:MutualExclusion2}.

~

\centerline{\epsfig{file=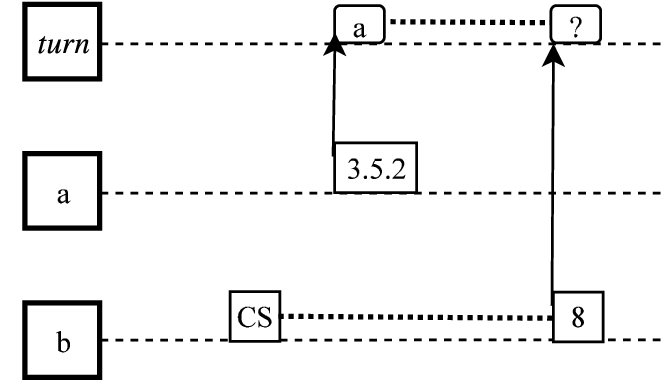,width=0.5\linewidth}}

\item Not at state $3.5.1$, because,
by lemma \ref{lemma:NoTwoIn351}, it can't be already
in $3.5.1$ and, as seen in the previous point,
there is no one in the extented critical section to
change $turn$ to FREE.

~

\centerline{\epsfig{file=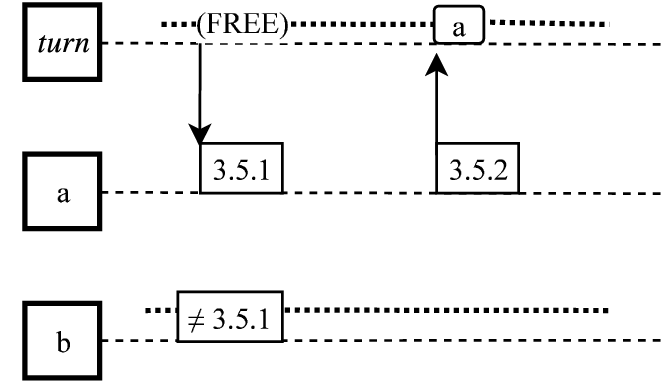,width=0.5\linewidth}}

\end{itemize}

If $turn$ is assigned to $a$ at state $8$ by a
process $b$, again, no other one may change this asigment:

\begin{itemize}
\item Not at state $3.5.2$.
For a contradiction, suppose
process $c$ is the first one in doing so.
$c$ must be at state $3.5.1$ ($turn=\text{FREE}$)
before $turn$ is set to $b$, so the assumptions for
this case hold.
If this can happen, then, $c$ could also reach
the critical section in $[\text{CS}_b, 8_b[$,
breaching theorem \ref{theorem:MutualExclusion2}.

~

\centerline{\epsfig{file=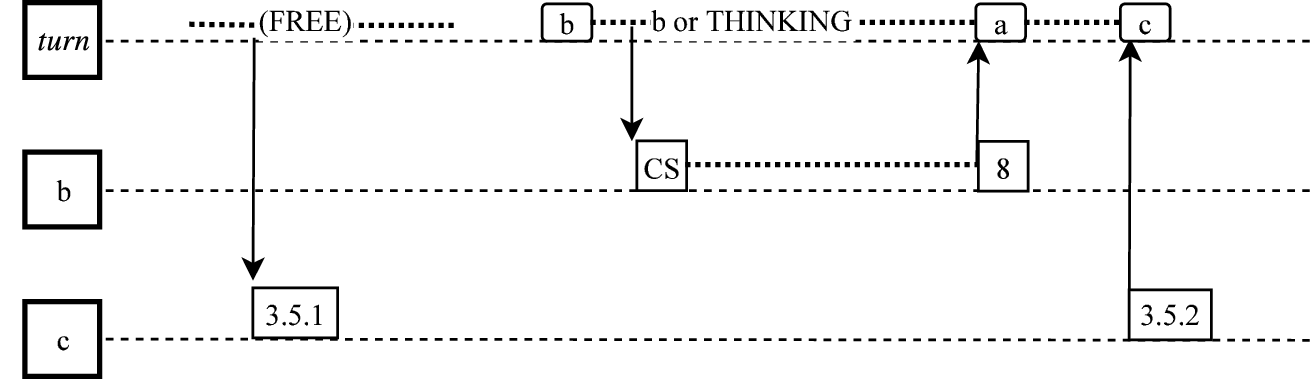,width=\linewidth}}

\item Not at state $8$. If, oppositely, a process $c$ is the
first to do this,
it must enter  the critical section after $8_b$ (not  to be overlapped
with $b$). But this would require a change in $turn$ to $c$ before 
the supposed first change after $turn=a$.

~

\centerline{\epsfig{file=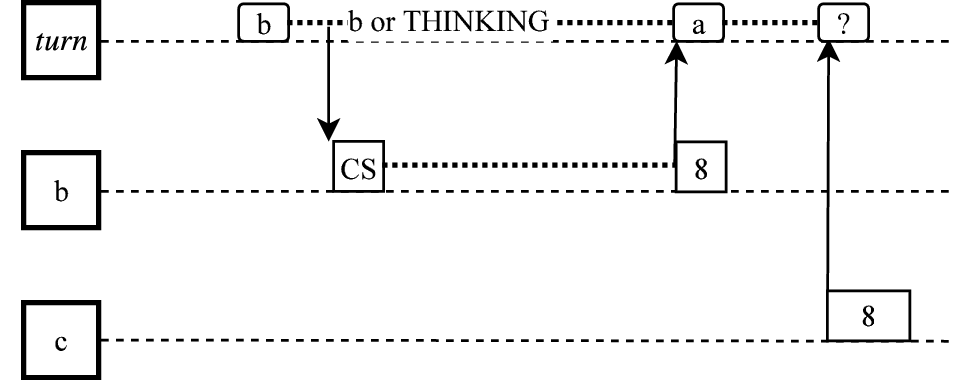,width=0.8\linewidth}}

\end{itemize}
$\Box$

\begin{lemma}
\label{lemma:ProcessWithTurnEnters2}

If the turn is assigned to a process, then, it will enter the critical
section.
\end{lemma}

{\bf Proof of lemma \ref{lemma:ProcessWithTurnEnters2}}

If a process sets $turn$ to itself
in $3.5.2$; it is able to reach
state $4$ --there is no condition against this.

A process in the extended critical section can't choose itself
for $turn$: it is not eligible  in state $6.1$ because its flag is
REMAINDER.

If $turn$ is assigned to a process in $8$, in case it is waiting
at state $2$, the assignment enables it
to get to state $4$.

Finally,
since an assignment of $turn$ to a process
doesn't change (lemma \ref{lemma:AssignmentFixed})
and it is able to arrive at state $4$, it will
eventually enter the critical section.
$\Box$

\begin{lemma}
\label{lemma:ElectTheMinimum}
A given process $a$ in $[5,8[$ will elect
another process $b$ for $turn$,
considering this order:  $a+1, ... , N-1, 0, 1, ... a-1$,
such that $b$
is smaller or equal
than the minimum identifier
of the processes
which {\em passed} state $2$ before process $a$ arrived at $5$.
\end{lemma}

{\bf Proof of lemma \ref{lemma:ElectTheMinimum}}

The loop defined by the states $6, 6.1, 6.2$ makes process $a$ to look
for a  waiting process  ($flag[j]\neq \text{REMAINDER}$)  with minimum
identifier following this  order: $a+1, ...  , N-1, 0,  1, ... , a-1$.
Every process that  has changed its flag to WAITING  in state 2 before
the loop begins (actually in state $5$) will be eligible for $turn$ in
state $6.1$.

There is the  corner case that a process $c$  becomes WAITING when the
loop has begun. In this case, $c$  will be considered only if the loop
counter  $j<c$.  Note  that $c$  will remain  in state  $2$ until  the
decision for $turn$ is made, because  $turn$ is set to THINKING by $a$
in $5$.  It will able to continue when $a$ reaches state $8$
and $turn$ is set.
$\Box$

\begin{corolary}
(of lemma \ref{lemma:ElectTheMinimum})
\label{corolary:FreeNo34}

When a process sets $turn$ to FREE, there is no process
in a state between $[3, 4]$.
\end{corolary}

{\bf Proof of corolary \ref{corolary:FreeNo34}}
Any process 
in a state between $[3, 4]$ has set its $flag[i]$ to
WAITING before $turn$ is set to THINKING by
the elector process in $5$.
Thus, it must be considered for $turn$ in $6.1$
This implies that $turn$ cannot be set to FREE.
$\Box$

\begin{lemma}
\label{lemma:IfFreeChooseOne}
If $turn$ is set to FREE, one of the waiting processes
will be elected for $turn$ in state $3.5.2$.
\end{lemma}

{\bf Proof of lemma \ref{lemma:IfFreeChooseOne}}

If $turn$ is set to FREE, by corolary
\ref{corolary:FreeNo34}, all waiting processes
must be at state $2$ before this assignment. The remaining processes
have $flag[i]=\text{REMAINDER}$.
The asignment to FREE allows processes waiting
at $2$ to continue ($turn\neq \text{THINKING}$).
At  least   one  of   them   will  reach   state  $3.2$  
(while $turn=\text{FREE}$), as well as at least one
will reach state $3.3$ (again while $turn$ is not set in $3.5.2$).

Let's consider the set of processes,  $S$, that reach state $3.3$ from
$3.1$ since $turn=FREE$ for the last time, and for which
their $flag[i]=$CANDIDATE.  $S$ is not empty, and
its size is bounded by $N$; this is, all the processes.

For a given execution, consider the
instant $t$ after which no more processes
are added to $S_t$ and let $a$ be the process with the smallest
identifier in $S_t$.

If  a  process  different  from  $a$ is  in  state  $3.5.1$  then,  by
lemma  \ref{lemma:NoTwoIn351},  $a$ won't  be  able  to reach  $3.5.1$
before that process sets $turn$ to  itself. And as a consequence, this
is proves the theorem for this case.

If on the contrary, no process is already in $3.5.1$,
$a$ will reach this state.

The  loop  defined by  states  $3.4,  3.4.1,  3.4.2$ counts  how  many
processes set $flag[i]=\text{CANDIDATE}$ in $3.2$
and which of them has the lowest identifier.

Any process $b \in S_t$, $a<b$, will reach state $3.5$ and because its
$minCandidate$ variable  is less  than $b$, it  will proceed  to state
$3.6$ and, finally, it will set $flag[b]$ to WAITING.  This means that
$b$ is no more a candidate, it will not reach state $3.3$ again in
this loop,  and that $b$ is  removed from $S_t$.  Eventually,  only $a$
will be  in $S_t$.   Then, $a$ will  find that  $turn=\text{FREE}$ and
$nCandidates=1$ (being  $minCandidate=a$).  This allows $a$  to assign
$turn$ to itself in state $3.5.2$.  $\Box$

\begin{theorem}
\label{theorem:ProcessWithTurnEnters2}
A waiting process ($flag[i]\neq REMAINDER$) will be elected for $turn$
and it will enter the critical section.
\end{theorem}

{\bf Proof of theorem \ref{theorem:ProcessWithTurnEnters2}}

By lemma
\ref{lemma:ElectTheMinimum},
a process in the extended critical section {\em fairly} chooses
the following process for $turn$ among waiting processes,
or sets it to FREE if no one is waiting when it performed
the election.

In     case      $turn$     is     FREE,     as      initially,     by
lemma \ref{lemma:IfFreeChooseOne},  $turn$ is eventually granted  to a
process.

By lemma \ref{lemma:AssignmentFixed}, once a process has the $turn$,
it is not changed before the process arrives at the critical section.
By lemma
\ref{lemma:ProcessWithTurnEnters2}, a process with the turn
will reach the critical section.

Finally, a process that sets its $flag[i]$ to WAITING, eventually arrives at
state $4$, and doesn't change this  setting, unless it reaches state $5$
after the critical section.

Therefore,  a waiting  process  will, eventually,  enter the  critical
section.

$\Box$


%
%

%

%


\bibliographystyle{plain}
\bibliography{tex/bibliography}

\end{document}